# AN OVERVIEW OF SEVEN ASTRONOMICAL DECADAL SURVEYS: Successes, Failures, and Shifting Demographics

Michelle Pan[1] and Virginia Trimble[2]

[1]*Independent Researcher*
[2]*Department of Physics and Astronomy, University of California, Irvine, CA, USA*

## ABSTRACT

We take a fresh look at about 60 years of recommendations for US federal funding for astronomical and astrophysical facilities provided by seven survey committees at roughly 10-year intervals. It remains true that very roughly one third of the highest priority items were done with (mostly) federal funding within about 15 years of the reports; another third happened with (mostly) state, private, or international funding; and about a third never happened (and we might well not want them now). Some other very productive facilities were never quite recommended but entered the queue in other ways. We also take brief looks at the long-term achievements of the highest-priority facilities that were actually funded and built more or less as described in the decadal reports. We end with a very brief look at the gender balance of the various panels and committees and mention some broader issues that came to look important while we were collecting the primary data. A second paper will look at what sorts of institutions the panel and committee members have come from over the years.

# 1 INTRODUCTION

Modern astronomy and astrophysics are astronomically expensive, and portions of the American community have been trying for more than 60 years to tell our sponsoring agencies how best to expend the available funds, which, of course, are never enough for everything everyone wants in the way of telescopes, satellites, auxiliary equipment, computing facilities, salaries, and all the rest. The telling has taken the form of decennial studies and reports thereon, starting in 1962, and frequently known by the surnames of the chairs of the main committees.

Thus we have had the Whitford Report, the Greenstein Report, Field, Bahcall, Taylor-McKee, Blandford, and most recently the Harrison-Kennicutt report, though the official names are different (Table I). Included are positions of the panel chairs at the time of the reviews, their connections with the American Astronomical Society, and h-indices at the present time for the papers on Astrophysics Data System that they wrote in the 30 years before the reports.

Section 2 updates the 2011 tables that summarized outcomes of the first four reports and outcomes and guesses at what would become of the fifth and six. Section 3 provides a very brief glimpse of the kind of information that will appear in Paper 2. Section 4 looks at some broader issues of competition for resources, implications for education, the interface between astronomers and other communities, and so forth.

Table 1.1 Panel Chairs for the Decadal Astronomical Reports, 1960s-2020s

| |
|---|
| Ground-Based Astronomy: A Ten-Year Program, 1964<br>Albert Whitford, Director, Lick Observatory<br>Whitford (h = 17)<br>AAS President 1967-70, Russell Lecturer 1986 |
| Astronomy and Astrophysics for the 1970s, 1972<br>Jesse Greenstein, Director, Palomar Observatory<br>Russell Lecturer (the first Jew) 1970<br>Greenstein (h = 44)<br>Defeated for AAS president 1974, VP 1955-57 |
| Astronomy and Astrophysics for the 1980s, 1982<br>George Field, Director Center for Astrophysics<br>Russell Lecturer 2014<br>Field (h = 33) |
| The Decade of Discovery in Astronomy and Astrophysics, 1991<br>John Bahcall, Prof. of Natural Sciences IAS |
| Warner Prize 1970, Russell Lecturer 1999<br>AAS President 1990-92<br>Bahcall (h = 67) |
| Astronomy and Astrophysics in the New Millennium, 2001<br>Christopher McKee, Prof. of Physics UCB (student of Field)<br>Russell Lecturer 2016, AAS Council 1981-83<br>McKee (h = 69)<br>Joseph Taylor, Prof. of Physics, Princeton<br>Nobel 1993, Heineman 1980<br>Taylor (h = 66) |
| New Worlds, New Horizons in Astronomy and Astrophysics, 2010<br>Roger Blandford, Prof. of Physics, Stanford<br>Blandford (h = 108)<br>Warner 1982, AAS Council 1999-2002, Heineman 1998 |
| Pathways to Discovery in Astronomy and Astrophysics for the 2020s, 2021<br>Fiona Harrison, Rosen Prof. of Physics, Caltech<br>AAS member<br>Harrison (h = 81)<br>Robert Kennicutt, Plumian Prof. Cambridge 2006-17, Prof. TAMU, U Az<br>Heineman 2007, ApJ Ed. 1999-2006<br>Kennicutt (h = 111) |

## 2 PRIORITIZED FACILITIES AND WHAT HAS BECOME OF THEM SO FAR

An earlier paper (Trimble 2011) tabulated the prioritized lists of desired facilities from the first five decadal surveys and the progress toward acquiring them, under the headings "What they asked for" and "What we got." The latter had five subheadings: (1) in operation within 15 years of the report, with mostly federal funding, (2) in operation within 15 years with mostly state, private, or foreign funding, (3) eventually built with mostly US federal funding, (4) eventually built with mostly other funding, and (5) not built up to 2010 and very unlikely for the still longer-range future. Of 106 identifiable requests in those five reports, about 1/3 each fell into categories (1), (2+3+4), and (5), and many of the (5)'s we would not have wanted by then, and would not now.

Section 2.3 updates the table for the 2001 report and adds in 2010 and hints for 2021. But first to be covered are a few overriding issues (not all pleasant ones) and a prejudiced look at very long time outcomes.

## 2.1 Are we all on the same side?

Astronomy of course has always had to compete with other sciences and their applications to sometimes urgent human needs and problems. The field is unlikely ever again to pull in as large a fraction of the GNP as it did during the lead up to the first Apollo moon landing in 1969 (which did yield spinoff benefits for ground-based astronomy), but over the years has done rather well.

There have, however, always been some fault-lines. Early on, there were two obvious issues for ground-based facilities, the balance to be achieved between optical and radio facilities and the priority given to optical facilities available to all American astronomers vs. those restricted to smaller groups having access to telescopes owned by individual universities and observatories. Space-based facilities were not prioritized until the Greenstein report for the 1970s.

Very roughly, this all played out as mostly public facilities for radio observing, continued support for some semi-private optical observatories, and space vs. ground firmly divided between NASA and NSF. The differences are, of course, often seen as inequities (sometimes with folks on both sides feeling badly treated).

To demonstrate, here are just three examples: (1) observing time granted on NASA-sponsored space observatories and whatnot has generally come with some money to process and publish the data, while observing time on either public or private optical telescopes has generally not come with automatic dollars, requiring the observer to propose separately for NSF support, (2) it has sometimes seemed like the division of NSF money between users and instrument builders for the public vs. private telescope strongly favored the former (which it also directly supported); this was a factor in at least one senior observer at the Hale (Mt. Wilson, Palomar) Observatories leaving for Europe, (Münch 1978), and (3) down to present institutional support and rewards in dollars and/or observing time provided to developers for focal plane instrumentation has not always been awarded equitably (Cohen 2022).

## 2.2 Were the Winners Worth Having?

Here are the peaks of what we got - the highest priorities of each of the first six surveys as eventually constructed, deployed, and used, ordered as optical, radio (or other wavelength band), and space. Our hope is that anyone likely to be reading this will recognize the eventual names of the facilities, so that it won't be necessary to say here just how much has been accomplished with them. The catch, however, is that we cannot do either a "placebo control" in which nothing of the sort was ever done or even a comparison with a "state of the art" in which something else with similar purpose was eventually built.

Table 2.1 Progress and current status of the major requests from the 1960s to 2000s

| |
|---|
| WHITFORD REPORT (for 1960s)<br>Three 150"-200" telescopes, one in the south, became two 150" at Kitt Peak and at Cerro Tololo.<br><br>Large Array of Pencil Beams operable down to 3cm, eventually became the Very Large Array or VLA |
| GREENSTEIN REPORT (for 1970s)<br>Test of multi element optical array built in 1979 as Multi Mirror Telescope, MMT, eventually replaced by single mirror in same mounting<br><br>VLA now under that name, on line 1978 (with subsequent major upgrades)<br><br>Four High Energy Astrophysical Observatories, of which three flown in 1970s, observing gamma rays, X-rays (Einstein), and cosmic rays |
| FIELD REPORT (for 1980s)<br>15 meter new technology telescope, reasonably approximated by Kecks I and II on line in 1990 and 1996<br><br>VLBA (Very Long Baseline Array) became operational in 1992<br><br>Advanced X-ray Facility in Space (AXAF) renamed Chandra after 1999 launch and is still operational as we write |
| BAHCALL REPORT (for 1990s)<br>Northern hemisphere 8m optimized for infrared is Gemini North (1999)<br><br>Millimeter Array started yielding early science in 2011 as ALMA (Atacama Large Millimeter and sub millimeter Array)<br><br>S (Shuttle) IRTF (infrared telescope facility) became S (Space) IRTF became Spitzer Space Telescope in 2003, one of the four Great Observatories<br><br>The top ranked moderate cost recommendation was Adaptive Optics, which indeed has gone forward and become essential |

> McKEE-TAYLOR REPORT (for 2000s)
> Giant Segmented Mirror Telescope (GSMT) will probably be achieved first by the Europeans and their E-ELT, European Extremely Large Telescope at ESO, with the American Giant Magellan (also in Chile) underway
>
> Expanded VLA, yes, with 2010 and 2020 request for Next Generation VLA as the fourth large scale priority
>
> NGST (New Generation Space Telescope, to cover UVOIR wavebands) somewhat shrunken in mirror diameter and descaled in wavelength coverage, etc., finally launched as JWST

Sharp eyes may notice some important facilities that were never quite prioritized, most conspicuously the COBE satellite (first to study the cosmic background radiation from space), which was a NASA-forced marriage of more than one response to a general announcement of opportunity.

More recently, the phrase multi-messenger astronomy flooded into the literature when LIGO detected its first binary neutron star merger just in time to be prioritized in the 2020 report.

**2.3 What They Asked For and What We Got**

Below, we have compiled some information regarding the recommendations and priorities offered by each decadal report from the 2000s to 2020s, and what has become of them. Those outcomes can be sorted into the same five categories used in 2011:

(1) in operation ≤ 15 years after report with mostly federal funding
(2) in operation ≤ 15 years after report with mostly other funding
(3) eventually built with mostly federal funding
(4) eventually built with mostly other funding
(5) never or very unlikely

Table 2.2 Progress and current status of large-small requests from 2000s McKee-Taylor report

| | |
|---|---|
| LARGE | |
| NGST (UVOIR) | Descoped/launched as JWST 2021 (3) |
| GSMT | HET (1998), SALT (2003), GMT, E-ELT in progress, TMT construction halted (2 and 4) |
| Athena | US withdrew from Con-X, merged with ESA |
| | & JAXA's IXO, now ESA's Athena (launch 2035?) (4) |
| Expanded VLA | Completed on time and within budget 2012 as Karl G. Jansky VLA (1) |
| LSST | In progress as Vera C. Rubin Observatory, first light 2024 (3) |
| Terrestrial Planet Finder (TPF) | Deferred indefinitely (5) |
| SAFIR (Single Aperture Far IR) | Not prioritized for 2010 or 2020 (5) |
| MEDIUM | |
| Telescope System Instrumentation Program | On-going, including instrument for Keck (1) |
| GLAST | Launched 2008 as Fermi Gamma-ray Space Telescope (1) |
| LISA (Laser Interferometer Space Antenna) | LISA Pathfinder in operation 2015-17, mostly ESA (4) |
| Advanced Technology Solar Telescope | Operational 2019 in Hawaii as Daniel K. Inouye Solar Telescope (3) |
| Square Kilometer Array (SKA) | MEERkat, HERA, ASKAP, MWA in operation 2022 as Pathfinders (Australia, So. Africa, UK) for 2027 first light (4) |
| Solar Dynamic Observatory (SDO) | NASA launch 2010, operates past design lifetime (1) |
| CARMA merger of OVRO & BIMA | Accomplished 2006; halted 2015 and land restored, millimeter observations now with ALMA (1) |
| EXIST (X-ray survey) | Not in 2010; dropped (5) |
| VERITAS (Cherenkov Array, UHE gammas) | In operation at Whipple Observatory from 2005 onward (1) |
| ARISE (Earth-space Radio Interferometry) | Canceled; US-Japan project also canceled (5) |
| FASR (Solar Array) | In operation, Owens Valley, from 2017 (3) |
| South Pole Submm Telescope | Rethought in 2018; now on hold with other Antarctic programs (2/4) |
| SMALL | |
| National Virtual Observatory | Operational 2007-14; archives remain (1) |
| LOFAR (Low Frequency Array) | Opened 2010, European, still expanding (2) |
| Advanced CR Comp. Experiment for Space Station (ACCESS) | AMS-01 on Space Station 2002, AMS-02 on SS 2011 (1) |
| Ultra-long Duration Balloon Flights | On-going; 100 hr achieved (1) |

The earlier conclusion remains that, up to this point in time, about 1/3 of the requests happened within 15 years and with mostly US federal funding; 1/3 happened later and/or with other funding; and 1/3 have never happened.

Moving on to the two most recent reports, we find that many of the recommendations are for astronomical or technical areas to be explored, with further competition for the realizations of these, and many final decisions left to the supporting agencies. Some of the specific recommendations are for facilities that had already appeared in earlier reports, so that adding them into the Scorecard table would not be fair.

Nevertheless, here are the Tables representing the Blandford (2010) and Harrison-Kennicutt (2020) prioritized items.

Table 2.3 Progress and current status of large-small requests from 2010s Blandford report

| | |
|---|---|
| LARGE | |
| WFIRST (Wide Field IR Space Telescope) | Going forward as Nancy Grace Roman Space Telescope (2027?) |
| Augmentation of Explorer Program | Imaging X-ray Polarimetry Explorer (SMEX-14) 2021, others planned |
| LISA | As above |
| IXO | As above |
| Vera C. Rubin Observatory | As above |
| Mid-Scale Innovations Program in Astronomical Sciences | Many funded, often technical development for bigger things, like ngVLA, BICEP, and IceCube-Gen2 |
| GSMT | As above |
| ACTA (Atacama Cherenkov Telescope Array) | CTA Observatory, Paranal & Canary Islands (11 countries + ESO not US) European 2022 highest ground-based priority |
| MEDIUM | |
| CCAT (Cornell-Caltech Atacama Telescope) | Became Cerro Chajnantor Atacama Telescope 25m, Descoped to 6m Fred Young Submm Ohs. US-Germany-Canada for 2024 |
| Exoplanet Program Science | Roman chronograph, star-shade etc. in progress |
| Inflation Probe | CMB items in progress |
| SMALL | |
| TSIP | As above |
| Gemini International Leadership | 6-nation collaboration still operated from NOIRLab |
| SPICA (Space IR Tel. for Cosmology and Astrophysics) | JAXA and ESA dropped in 2020 |
| TCAN (Theoretical and Computational Astrophysics Networks) | Competitions in 2011, 2017, 2020, 2022, with rule changes |
| TOTAL # of Requests | 15 |

The most recent decadal report includes 20 recommended items, more or less prioritized. Many were already represented in earlier reports; some others are generic ideas; and three are for US involvement in projects where investment by other countries will dominate.

Table 2.4 Progress and current status of large-small requests of 2020s Harrison-Kennicutt report

| | |
|---|---|
| LARGE/MEDIUM | |
| LUVOIR | Highest priority space mission, 2029(?), First of new Great Observatories Program; roughly what NGST was intended to be |
| US ELT | As above, as GSMT (2028?) |
| ngVLA | Follow-on to VLA & ALMA |
| CMB-S4 | Ground-based CMB, Atacama & South Pole, NSF + DOE, South Pole site problems |
| IceCube-Gen 2 | Technical development in progress |
| JWST | As above; was NGST (UVOIR), launched 2021 |
| Roman Space Telescope | As above, was WFIRST, 2027(?) |
| US Involvement in Euclid | Launched 1 July 2023; led by ESA |
| US Involvement in Athena (Advanced Telescope for High Energy Astrophysics) | Began as Con-X, then IXO, now mostly ESA (2024?) |
| US Involvement in LISA | As above, Pathfinder successful, ESA and others (mid 2030s?) |
| Time Domain and Multi-Messenger Program (TDAMM) | Highly prioritized, separate from space frontier missions, in progress w/ significant $$$ |
| MEDIUM/SMALL | |
| NASA Astrophysics Research and Analysis | In progress, should focus on larger projects |
| Strategic Astrophysics Technology | Carry-over from 2010, to be continued |
| NSF Advanced Technologies and Instrumentation | Also very broad; accepting proposals |
| Suborbital balloons and sounding rockets | Aim for more, higher, longer duration, etc. |
| Explorer Program Augmentation | As above; carry one with SMEX, MIDEX missions |
| Pioneers Program (support for early career researchers in space instrumentation) | In progress |
| NSF MRIP | As above |
| TOTAL # of Requests | 20 |

We note that, as ever, the folks who actually invent, design, and build the entities we all need turn up fairly far down the priority list; and also that some expansion of the Antarctic neutrino observatory was supposed to be in here somewhere (as the last of about 5 expensive, prioritized items, but now also temporarily waylaid by need to dredge things out from snow accumulations and to provide more electric power at the active sites.

Additionally there has been quite a lot of rearrangements in the prioritization of requests. Several of the requests from the 2000s report have "moved up" in priority and endorsement, whereas several have been dropped. For instance, the Vera C. Rubin Observatory became the top-ranked priority in the 2010s report, but the Con-X mission from the 2000s report, which was once ranked the highest priority, was ranked fourth in this survey. Clearly the reorganization of priority reflects the development and progress in astronomy and astrophysics research and development in the past decades.

In the recent two decades we begin to see the emergence of astrobiology. Much emphasis is placed on the concept of life beyond Earth–both its evolution and past, and its future potential. Though the study of astrobiology first began in the mid-20th century, it seems to be on the rise nowadays, in the 2010s and 2020s.

Table 2.5 Requests Scorecard

| Report | Number of identifiable items requested | Number in operation ≤ 15 years after report w/ mostly federal | Number in operation ≤ 15 years after report w/ mostly other funding (state, | Number eventually built w/ mostly federal funding | Number eventually built w/ mostly other funding | Never or very unlikely |
|---|---|---|---|---|---|---|

|  |  | funding | private, foreign) |  |  |  |
|---|---|---|---|---|---|---|
| Whitford | 13 | 6 | 0 | 0 | 1 | 6 |
| Greenstein | 21 | 5 (+1 similar) | 2 | 4 | 1 | 8 |
| Field | 21 | 3 (+2 similar) | 2 | 3 | 2 | 9 |
| Bahcall | 29 | 11 | 6 | 5 | 0 | 7 |
| McKee - Taylor | 23 | 8 | 1 | 5 | 3 | 6 |
| Blandford | 23 | 10 | 1.5 | 4 | 3.5 | 4 |
| **Total** | **129** | **45 (+3 similar)** | **13** | **20** | **11** | **38** |

**2.4 NASA Launches**

Table 2.6 2010s NASA Launches (pertaining to astronomy and the Solar System) (List of NASA Missions, Wikipedia)

| Mission Name | Start/Launch Date |
|---|---|
| **Astronomy/physics Missions** | |
| NuSTAR | 2012, operational |
| Parker Solar Probe | 2018, operational |
| Solar Dynamics Observatory | 2010, operational |
| **Solar System Missions** | |
| Discovery 12 – InSight | 2018, completed |
| MAVEN under Mars Scout Program | 2013, operational |
| MSL Curiosity Rover | 2011, operational |
| New Frontiers 2 – Juno | 2011, operational |
| New Frontiers 3 – OSIRIS-REx | 2016, operational |

Table 2.7 2020s NASA Launches (pertaining to astronomy and the Solar System) (List of NASA Missions, Wikipedia)

| Mission Name | Start/Launch Date |
|---|---|
| **Astronomy/physics Missions** | |
| James Webb Space Telescope | Launched 2021, operational |
| Nancy Grace Roman Space Telescope | Planned launch in 2027 |
| **Solar System Missions** | |
| Mars 2020: Perseverance | Launched 2020, operational |
| Mars 2020: Ingenuity | Launched 2020, operational |
| SHIELDS | Launched 2021 |
| Discovery 13 – Lucy | Launched 2021, operational |
| Double Asteroid Redirection Test | Launched 2021, done |
| CubeSat for Solar Particles (CuSP) | Launched 2022, done |
| Discovery 14 – Psyche | Planned launch in 2023 |
| EscaPADE | Planned launch in 2024 |
| Europa Clipper | Planned launch in 2024 |
| Solar Terrestrial Probes Program: IMAP | Planned launch in 2025 |
| New Frontiers 4 – Dragonfly | Planned launch in 2027 |
| Discovery 15 – VERITAS | Planned launch in 2028 |
| Discovery 16 – DAVINCI | Planned launch in 2029 |

## 3 DEMOGRAPHICS

As the times have changed throughout the decades (from the 1960s to the 2000s), it appears that the participation of women on the committees and sub-panels for the decadal reports has increased. The growth of the participation of women in a male-dominated field is a first step towards breaking gender stereotypes and promoting gender diversity. Additionally, we have seen notable changes with regards to the types of institutions present in such committees in the decadal reports. In the 1960s, individuals representing independent observatories seemed to make up the majority of said population, whereas members of research universities and national laboratories or observatories now prominently take the lead by far.

### 3.1 Gender Ratios in Recent Steering Committees

Women were barely present in the committees in the earlier decadal reports (Trimble 2011), but are now much more represented than before, making up as much as half of the panel population:

Table 3.1

Steering Committee for the 2010s Decadal Survey of Astronomy and Astrophysics

| | | |
|---|---|---|
| **Total** | 23 | 100% |
| Male | 17 | 74% |
| Female | 6 | 26% |

Table 3.2

Steering Committee for the 2020s Decadal Survey of Astronomy and Astrophysics

| | | |
|---|---|---|
| **Total** | 20 | 100% |
| Male | 12 | 60% |
| Female | 8 | 40% |

    As can be seen, the representation of men vs. that of women is still, on average, greater for the former and lesser for the latter. However, the change from the first decadal report is tremendous (Trimble 2011), as more and more women are entering the field. It appears that promoting access to educational opportunities in this field is extremely impactful especially for women and minorities who typically face obstacles that limit their ability to acquire such options. As such, it has become more plausible to facilitate an environment that enables more diversity in both gender and ideas.

    Paper 2 will examine gender distributions for all the panels and take a look at the home institutions of all the Committee and Panel members engaged in the 1990 (Bahcall), 2010 (Blandford), and 2020 (Harrison-Kennicutt) decadal surveys. We were a bit surprised by some of the details.

**4 CONCLUDING ANALYSIS**

    Overall it seems as though many programs and missions recommended or prioritized by the decadal reports have been canceled or deferred due to budgetary and funding concerns. U.S. missions and plans seem to experience a lot of hindering as a result of financial and infrastructural issues. Though this issue seems basic, it is one that is fundamental to the very progress and development of the elaborate plans themselves. Many of the funding programs that were mentioned in the 2000s and 2010s reports were not prioritized; however, it seems that the

TSIP program and a couple others that were prioritized have been helpful in facilitating more developments and progress in technology and instrumentation. Later, these programs seem to have been prioritized more and it appears that the suggested ones such as ATI and MRIP may be helpful for this situation. As these programs are increasingly more open to the public, rather than specific groups of individuals, the community is able to receive more input and diverse ideas from a wider audience.

However, the question of when these programs will be implemented and how much impact will result, are to be considered. Many of the programs are for entry to mid-level researchers to engage in a project as a PI; while that is undoubtedly helpful especially financially, how much these programs will aid the infrastructural issues (for instance, the CMB-S4 situation at the South Pole with the lack of electricity) is worth questioning.

The causes of the prevailing budgetary issues often come from things like the sunk-cost fallacy. The final costs of many, many projects end up summing up to much more than initially proposed during the planning stage. It is unclear whether the reason behind this is due to a lowballed initial number to get the project started, more expenditure of resources instead of abandonment due to the investment already underway (though of course project deferrals do also take place), or a combination of both.

The cancellation, deferral, or, at least, hindrance of the execution of projects appears to be a trend that has been prevalent for decades. Perhaps it may be a good idea to continue to look further into such program initiatives (and their potential variations) and take them into more consideration for future works.

Even when a project does not experience indefinite deferral or some sort of hindrance, however, it seems to be a trend that it is to take at the very least a decade to realize itself. The amount of time elapsed between the initial request and the actual date of construction or launch has grown from less than 10 years to much longer. Projects discussed more than a decade ago are finally being built or launched now, or have a completion date scheduled in the (near) future.

Meanwhile, the issue of finding an appropriate location to facilitate the construction of ground-based observatories proves to be a significant matter. It is important to, of course, consider the nature and quality of the data that can be collected depending on the location. However, in addition to the scientific perspective, consideration of the environment, culture, and heritage behind the chosen locations continues to be essential. For instance, the construction of

TMT at Maunakea, Hawai'i was already underway, but due to its controversial location which conflicts with the beliefs and heritage of some people whose ancestors lived there before invaders colonized the lands, progress has been halted. Hence certain spots may be culturally significant to the ideals of important groups of people, and this is also something to take into consideration and look into for future selection of ground-based facilities.

Ultimately a foundational matter for the growth of the astronomy and astrophysics community is the access to educational opportunities for the public. Promoting education in the sciences, particularly in astronomy and physics, is truly valuable. Over time, this will have the greatest positive impact as encouragement of, support for, and access to such education can help expand the community's knowledge base in search for more innovative ideas. It will also facilitate faster development, growth, and physical construction of both ongoing and future projects, as well as collaboration between different scientists, with whom the exchange of knowledge is invaluable.

**ACKNOWLEDGEMENTS**

We would like to acknowledge the usage of the "New Worlds, New Horizons in Astronomy and Astrophysics" Decadal Survey and the "Astro2020 Decadal Survey: Priorities for Small and Mid-Scale Projects" in our paper.